\documentstyle[prd,aps,twocolumn,eqsecnum,epsf]{revtex}

\begin{document}


\draft
\widetext

\title{Quantum Formation of Black Hole and Wormhole \\in Gravitational
  Collapse of a Dust Shell}

\author{Kouji Nakamura\thanks{e-mail address:
    kouchan@allegro.phys.nagoya-u.ac.jp}, Yoshimi
  Oshiro\thanks{e-mail address: yoshiro@allegro.phys.nagoya-u.ac.jp}
  and Akira Tomimatsu\thanks{e-mail address:
    c42615a@nucc.cc.nagoya-u.ac.jp}}

\address{Department of Physics, Nagoya University, Chikusa-ku, Nagoya
  464-01, Japan}

\date{June 10, 1995}

\maketitle

\begin{abstract}
  Quantum-mechanical model of self-gravitating dust shell is
  considered. To clarify the relation between classical and quantum
  spacetime which the shell collapse form, we consider various time
  slicing on which quantum mechanics is developed. By considering the
  static time slicing which corresponds to an observer at a constant
  circumference radius, we obtain the wave functions of the shell
  motion and the discrete mass spectra which specify the global
  structures of spherically symmetric spacetime formed by the shell
  collapse. It is found that wormhole states are forbidden when the
  rest mass is comparable with Plank mass scale due to the zero-point
  quantum fluctuations.
\end{abstract}

\section{Introduction}
  \label{sec:intro}

        Historically, many works have been devoted to constructing
        quantum theories of gravitational collapse. One of the
        motivations is to resolve the problems concerning the final
        fates of black hole evaporation due to the Hawking
        radiation\cite{Hawking-rad}, which may lead to the paradox of
        information loss\cite{infloss}. However, quantum gravity
        itself includes not only technical difficulties but also
        conceptual ones such as the interpretation of wave function,
        the nature of time and the definition of observables. Since
        any fully consistent theory is not yet established, the
        present step would be to develop various useful toy models to
        shed a new light on some features of quantum effects of
        gravity. In particular, quantum gravitational collapse of a
        spherically symmetric shell, on which we focus our attention
        in this paper, has been studied as one of such toy
        models\cite{BKKT,Berezin90,iroiro,Hajicek-math,Hajicek-phys}.

        In this model, one can consider the spherically symmetric
        spacetime which is in vacuum except the delta-function
        distribution of incoherent dust shell at a finite
        circumference radius. By the virtue of the spherical symmetry,
        the inner side of the shell may be a Minkowski spacetime and
        the outer one is the Schwarzschild spacetime which has a
        finite gravitational mass parameter $M$. The divergence of the
        energy momentum tensor of the dust shell tells us the shell
        can be characterized by a constant of motion which are called
        the rest mass $\mu$. The two vacuum regions should be joined
        at the shell according to the Einstein equations. This program
        can be accomplished by using Israel's junction
        condition\cite{Israel}, which works as the equation of motion
        of the dust shell, and the dynamical variable is limited to
        the circumference radius $R$ of the shell and the global
        structure of spacetime can be classified by the value of $E
        \equiv M/\mu$.

        The original idea to quantize this dynamical system was
        proposed by Berezin et al.\cite{BKKT,Berezin90}. If any quantum
        effect of incoherent dust shell is ignored, this matter shell
        can be characterized by $\mu$. Both the vacuum regions in the
        spacetime are also classically treated. Then, only the
        equation of motion of the shell which can be regarded as the
        energy equation is quantized by constructing the Hamiltonian
        operator. Although the Hamiltonian operator to be constructed
        is not unique, Berezin et al. investigated the Hamiltonian
        operator,
    \begin{equation}
      \label{B-h}
      H_{B} = \mu \cosh \left( - \frac{i\hbar}{\mu}
      \frac{\partial}{\partial R}\right) - \frac{m^2}{R},
    \end{equation}
        where $m = \mu /m_{p}$ and $m_{p}$ is the Planck mass
        (Throughout this paper we denote the Planck constant by
        $\hbar$ and use units such that $G = c = 1$) and the time
        coordinate describing the shell dynamics is chosen to be the
        proper time of a comoving observer. Then, they obtained the
        spectrum of the eigenvalues $E$ of $H_{B}$,
    \begin{equation}
      \label{berezin-e}
      E = 1 - \frac{m^4}{8(n+1)^2},
    \end{equation}
        by using WKB approximation where $n$ is a non-negative
        integer. From their analysis, it is unclear that the spectrum
        (\ref{berezin-e}) is also valid when $m^2>>n$. However,
        H\'aj\'\i\v{c}ek et al.\cite{Hajicek-phys} gave a definite
        formula of the spectrum $E$ adapting a simpler equation for
        the wave function. They introduce the super-Hamiltonian on an
        extended minisuperspace which leads to the ``Wheeler-DeWitt''
        equation,
    \begin{equation}
      \label{Hajicek-WD}
      \left(- i\hbar \frac{\partial}{\partial T} - \frac{m^2}{2R}
    \right)^{2} \Psi + \frac{\partial^{2}}{\partial R^{2}}\Psi - \mu^2
    \Psi = 0,
    \end{equation}
        where $T$ is the Minkowskian time of the inner side of the
        shell. Investigating this equation, they obtain the formula of
        $E$ for the bound states defined by $-1<E<1$,
    \begin{equation}
      \label{comoving-eigen}
      E = \frac{2(\kappa + n)}{\sqrt{m^4 + 4(\kappa + n)^2}},
      \;\;\;\;\;\; \kappa = \frac{1}{2} + \frac{1}{2}\sqrt{1 - m^4}.
    \end{equation}
        Nevertheless the two quantum treatments arrive at the same
        conclusion when $m^2<<n$, (\ref{comoving-eigen}) becomes
        meaningless when $\mu > m_{p}$ which means classical limit
        cannot be obtained from this eigenvalue. Furthermore, it is
        shown that in the bound states only the black hole formation
        corresponding to the range $1/2<E<1$ is allowed, while the
        wormhole formation corresponding to $0<E<1/2$ is possible as
        classical solutions.

        Our purpose in this paper is to clarify the relation between
        quantum and classical spacetime of this system. For this
        purpose, we pay attention to the time slicing on which the
        quantum mechanics is developed. Note that the previous
        treatments are essentially based on the comoving time slicing.
        It is also possible to construct quantum mechanics on the
        other time slicing. Since the canonical formalism is based on
        the decomposition of spacetime into space and time, a special
        attention must be paid to the problem of time slicing which
        determines the foliation of spacetime. Furthermore it is
        well-known in classical relativity how a description of black
        hole spacetime depends on the choice of time slicing: In the
        usual static chart the Schwarzschild horizon plays a role of
        the infinite redshift surface, while in the synchronous chart
        corresponding to a freely falling observer it is not any
        special surface\cite{Landau}. This means that the horizon can
        be regarded as a sort of boundary of the foliated spacetime
        only for a static observer. So the foliation of spacetime or
        the choice of observers in the spacetime is more important
        when one consider a black hole spacetime.

        To study the quantum mechanics of dust shell collapse, we use
        various time slicings in hope that the physical essence is
        independent of the time slicing. The black hole horizon is not
        a special surface for an observer who uses the proper time
        along the shell history or the Minkowskian time. However, a
        static observer outside the horizon will require a different
        boundary condition for the wave function due to the existence
        of the horizon for him. By developing quantum mechanics for a
        static obsrever, we can obtain the mass spectrum which
        corresponds to (\ref{berezin-e}) and
        (\ref{comoving-eigen}). Furthermore, we will show that
        wormhole states are also possible for a static observer who
        stays inside the wormhole but when $\mu$ is same order of
        $m_{p}$ no wormhole state is allowed owing to the zero-point
        fluctuation of the shell motion.

        This paper is organized as follows. In
        Sec.\ref{sec:eqofmotion}, we derive the classical equation of
        motion of the shell in terms of various observers who define
        time slicings on the shell trajectory in spacetime. Although
        infinitely many observers or time slicings can be assumed in
        general, we restrict our consideration to a one-parameter
        family which contains two typical time slicings corresponding
        to the Gaussian normal coordinates of a comoving observer and
        the static Schwarzschild coordinates of an observer who stays
        at a finite circumference radius, and derive the possible
        spacetimes as the solutions of the Einstein equations. In
        Sec.\ref{sec:canqua}, we introduce the Hamiltonian constraint
        which corresponds to the time-time component of the Einstein
        equations and gives the equation of motion in a similar manner
        to the Wheeler-DeWitt procedure. Following the earlier works,
        we solve classically the Einstein equations except the
        Hamiltonian constraint for the shell motion. This means that
        the momentum constraint is classically treated just like many
        minisuperspace models.

        In Sec.\ref{sec:eigen}, we consider the quantum mechanics of
        collapsing shell, by using the one-parameter family of time
        slicings introduced in the previous sections and we also
        discuss the global structure of spacetime in which the dust
        shell collapse forms. Our main result will be obtained under
        the static time slicing. Furthermore, our consideration is
        restricted to the so-called bound states $ -1 < E < 1$, which
        have the discrete mass eigenvalues, and we must consider the
        cases $E>1/2$ and $E<1/2$ separately in the static time
        slicing, these cases correspond to the black hole states and
        wormhole states respectively. Then, we discuss the relation
        between quantum and classical solutions of this system and
        show that there is no quantum states when $E<1/2$ and $\mu\sim
        m_{p}$. We also consider the quantum mechanics on non-static
        time slicing to confirm that our arguments are natural
        extension of the result obtained in the comoving frame. It can
        be shown by the fact that the quantum version of the
        Hamiltonian constraint becomes identical with the radial
        equation of (\ref{Hajicek-WD}) in the comoving limit. Finally,
        our consideration was summarized in Sec.\ref{sec:sum}.
        Although our simple model is a preliminary approach to quantum
        gravity, it would give a useful clue when one investigates the
        problem of time slicing in a more complete theory.

\section{classical equation of motion}
        \label{sec:eqofmotion}

        In this section, we derive the equation of motion for a
        collapsing dust shell, which is described by various time
        slicing. Since the spacetime is spherically symmetric, one can
        choose the metric of the form
    \begin{equation}
      \label{eqn:general-spherically}
        ds^2 = g_{ab}dx^{a}dx^{b} + R^2 d\Omega^{2}_{2}, \;\;\;\; (a,b
        = 0,1)
    \end{equation}
        where $R$ is the circumference radius and $d\Omega^{2}_{2}$ is
        the metric of unit 2-sphere. For this metric matter fields
        depend on $x^{0}$ and $x^{1}$ which are time and radial
        coordinates.  Furthermore the spacetime is assumed to be in
        vacuum except the $\delta$-function matter distribution at a
        finite circumference radius. We denote the world volume of
        this spherically symmetric shell as $\Sigma$ which is a
        (1+2)-dimensional hypersurface. In a neighborhood of $\Sigma$,
        one can choose the Gaussian normal coordinates\cite{Blau},
    \begin{equation}
      \label{eqn:comoving-spherically}
        ds^2 = -d\tau^2 + d\eta^2 + R^2 d\Omega^{2}_{2},
    \end{equation}
        where $\tau$ is the proper time which would be measured by an
        observer comoving with the shell. The coordinate $\eta$ is the
        proper distance from $\Sigma$ along the geodesics which are
        orthogonal to $\Sigma$, and $\Sigma$ is assigned to the world
        volume of $\eta=0$. Imposing the spacetime to be orientable on
        $\Sigma$, we may regard one side of $\Sigma$ as being the
        ``outer side'' ($\eta>0$) and the other side as being ``inner
        side'' ($\eta<0$).

        The whole spacetime is constructed by connecting the
        Schwarzschild spacetimes of the outer side and the Minkowski
        spacetime of the inner side. The Einstein equations give the
        ``junction condition'' for the neighborhood of $\Sigma$, which
        is well-known as Israel's formula\cite{Israel,Blau}. In the
        case of the Minkowski-Schwarzschild junction, the nontrivial
        conditions are given by\cite{Blau},
    \begin{equation}
      [\partial_{\eta}R] = - \frac{\mu}{R},\;\;\;\;\;\;
      [\partial_{\tau}A] = 0,
          \label{nontrivial-junction}
    \end{equation}
        where $\partial$ denotes the partial derivative with
        respect to its subscripted coordinate, and $A$ represents
        all the metric functions on the spacetime. The bracket $[A]$
        means the difference of $A$ between the outer and inner sides,
    \begin{equation}
       [A]=\lim_{\eta \rightarrow +0}A - \lim_{\eta \rightarrow -0}A.
    \end{equation}
        The first equation of (\ref{nontrivial-junction}) contains the
        total energy of dust shell defined by $\mu = \sigma R^2$ where
        $\sigma$ is the surface energy density of the dust shell. From
        the divergence of the energy-momentum tensor of dust shell,
        one can easily see that $\mu$ is a constant of motion.

        For our convention of calculation, let us introduce quasi-local
        mass defined by
    \begin{equation}
      M = \frac{R}{2}(1-g^{\mu \nu}\partial_{\mu}R\partial_{\nu}R),
          \label{eqn:hawking-mass}
    \end{equation}
        which is conserved in each vacuum region of the spherically
        symmetric spacetime. Of course, $M=0$ in the
        Minkowski side, and $M(\neq0)$ in the Schwarzschild side
        represents the gravitational mass of the shell. The formula
        (\ref{eqn:hawking-mass}) is useful to estimate the
        derivatives of $R$ in both sides, which are involved in
        (\ref{nontrivial-junction}).

        Now we discuss the time slicings to describe the shell
        motion. ``Time slicings'' usually mean foliations of a whole
        spacetime. Foliations are spaces in a spacetime in which one
        can set observers. In our model, however, the equation of
        motion for the collapsing shell can be reduced to a local
        equation on $\Sigma$. Hence, the necessary procedure is to set
        a radial direction of coordinate system near $\Sigma$, which
        is called ``time slicing'' in this paper. Let us denote the
        radial coordinate by $x$ and rewrite the metric into the form
    \begin{equation}
      \label{semi-general-metric}
      ds^{2} = - N^{2}dt^2 + U^2 dx^2 + R^2 d\Omega^{2}_{2}.
    \end{equation}
        We can refer to a local observer near $\Sigma$ whose world
        line is along a constant $x$. There are two typical
        observers. One is a comoving observer corresponding to the time
        slicing $x=\eta$. The world line of this observer is embedded
        in $\Sigma$. Another is a static observer who stays at a
        constant circumference radius $x = R$ ($dx = dR$ in
        (\ref{semi-general-metric})). We call the above time slicings
        ``comoving slicing'' and ``static slicing,'' respectively. Our
        idea is to give a more general form of the time slicing as
        follows,
    \begin{equation}
      \label{eqn:gauge-choice}
      dx = \xi dR + \zeta d\eta,
    \end{equation}
        where $\xi$ and $\zeta$ are constant parameters at the outside
        the shell. This form is useful because one can recover the
        static and comoving time slicings by choosing the parameters
        to be $\zeta = 0$, and $\xi=0$, respectively. Except the
        special cases the time slicing corresponds to an observer who
        is not comoving with the shell but infalling toward the origin
        $R=0$. In the following we will study the parameter dependence
        of the equation of the shell motion in both classical and
        quantum levels.

        The first task is to rewrite the junction condition
        (\ref{nontrivial-junction}) using the coordinates $x$ and
        $t$. The general coordinate transformations from
        (\ref{eqn:comoving-spherically}) into
        (\ref{semi-general-metric}) should be
    \begin{eqnarray}
      \label{coordinate-trans-1}
      d\tau&=&N\cosh\phi dt + U\sinh\phi dx,\\
      d\eta&=&N\sinh\phi dt + U\cosh\phi dx.
      \label{coordinate-trans-2}
    \end{eqnarray}
        The boost angle $\phi$ is related to the velocity of the shell
        in the $(t,x)$ frame as follows
    \begin{equation}
      \label{eqn:boost-angle}
      \tanh\phi = - \frac{U}{N}\left(\frac{dx}{dt}\right)_{\eta}.
    \end{equation}
        where the subscript $\eta$ means the derivative along a line
        of constant $\eta$. In this frame the junction conditions
        (\ref{nontrivial-junction}) are given by
    \begin{equation}
      \label{gene-junc}
      [(\partial_{x}R)_{t}] = - \frac{\mu}{R} U \cosh\phi,\;\;\;\;\;\;
      [(\partial_{t}R)_{x}] = - \frac{\mu N}{R}\sinh\phi,
    \end{equation}
        where $t$ and $x$ are treated as independent variables in the
        calculation of the partial derivatives.

        On the other hand, because the quasi-local mass $M$ defined by
        (\ref{eqn:hawking-mass}) invariant under the coordinate
        transformation (\ref{coordinate-trans-1}) and
        (\ref{coordinate-trans-2}), it must satisfy with the
        conditions
    \begin{equation}
      \label{general-mass-p}
      \frac{2M}{R} - 1 = \frac{(\partial_{t}R)_{x+}^{2}}{N^2} -
      \frac{(\partial_{x}R)_{t+}^{2}}{U^2},\\
    \end{equation}
        in the Schwarzschild side, and
    \begin{equation}
      \label{general-mass-n}
      - 1 = \frac{(\partial_{t}R)_{x-}^{2}}{N^2} -
      \frac{(\partial_{x}R)_{t-}^{2}}{U^2},
    \end{equation}
        in the Minkowski side. By the virtue of the time slicing
        (\ref{eqn:gauge-choice}) and the coordinate transformations
        (\ref{coordinate-trans-1}) and (\ref{coordinate-trans-2}), we
        obtain the relations
    \begin{eqnarray}
      \label{rela-dev-1}
      \left(\frac{dx}{dt}\right)_{\eta\pm} &=& \xi
      \left(\frac{dR}{dt}\right)_{\eta\pm}, \\
      \label{rela-dev-2}
      \left(\frac{dR}{dt}\right)_{x\pm} &=& -
      \frac{\zeta}{\xi}\left(\frac{d\eta}{dt}\right)_{x\pm} = -
      \frac{\zeta}{\xi}N\sinh\phi, \\
      \label{rela-dev-3}
      \left(\frac{dR}{dx}\right)_{t\pm} &=& \frac{1}{\xi} -
      \frac{\zeta}{\xi}\left(\frac{d\eta}{dx}\right)_{t\pm} =
      \frac{1}{\xi} - \frac{\zeta}{\xi}U\cosh\phi.
    \end{eqnarray}
        If the metric components $N^2$ and $U^2$ in the coordinate
        system $(t,x)$ are assumed to change continuously at the
        shell, substitution of (\ref{rela-dev-2}) and
        (\ref{rela-dev-3}) into (\ref{gene-junc}) leads to
    \begin{equation}
      \label{eqn:finaljunction}
      [\zeta] = \frac{\mu\xi}{R},\;\;\;\;\;\; [\xi] = 0.
    \end{equation}
        Note that the parameter $\zeta=\zeta_{+}$ at the outer side
        must be different from $\zeta=\zeta_{-}$ at the inner
        side. Since we are interested in an observer located at the
        outer side of the shell, only $\zeta_{+}$ is treated as a
        parameter for specifying the time slicing. Then $\zeta_{-}$
        becomes a function depending on $R$. Furthermore,
        (\ref{general-mass-p}) and (\ref{general-mass-n}) for the
        quasi-local mass in both sides are rewritten into the
        forms
    \begin{equation}
         \label{eqn:outmassver2}
      \left(\frac{2M}{R}-1\right)\xi^2 = \frac{2\zeta_{+}\cosh\phi}{U}
      - \zeta_{+}^2 - \frac{1}{U^2},\\
    \end{equation}
        and
    \begin{equation}
        \label{eqn:inmassver2}
      - \xi^2 = \frac{2\zeta_{-}\cosh\phi}{U} - \zeta_{-}^2 -
      \frac{1}{U^2},
    \end{equation}
        respectively. If $\zeta_{-}$ and $\phi$ are omitted from
        Eq.(\ref{eqn:finaljunction}), (\ref{eqn:outmassver2}) and
        (\ref{eqn:inmassver2}), we obtain the metric component
    \begin{equation}
      \label{u-inverse}
      \frac{1}{U^2} = \zeta^2_{+}+ \left(1 - \frac{2M}{R}\right)\xi^2
      + \zeta_{+}\xi\left(\frac{2M}{\mu}-\frac{\mu}{R}\right),
    \end{equation}
        which shows that $g_{xx}=1$ in the comoving time slicing ($\xi
        = 0, \zeta_{+} = 1$), while $g_{xx}=1/(1-2M/R)$ in the static
        time slicing ($\xi = 1, \zeta_{+} = 0$). The velocity of the
        shell motion measured by the time coordinate $t$ should be
        defined by $(dR/dt)_{\eta}$. Then, by using
        (\ref{eqn:boost-angle}), (\ref{rela-dev-1}),
        (\ref{eqn:outmassver2}) and (\ref{u-inverse}), we arrive at
        the final form of the equation of motion
    \begin{equation}
        \label{general-equation-of-motion}
      \frac{\mu}{2N^{2}} \left(\frac{dR}{dt}\right)^2_{\eta} +
      \frac{1}{2\mu} V(M,\mu,\lambda,R) = 0,
    \end{equation}
        where the potential $V = V(M,\mu,\lambda,R)$ is given by
    \begin{eqnarray}
      V &=& V_{\infty}\left(1 + \frac{\displaystyle
        V_{\infty}}{\displaystyle \mu^2\left[\lambda
          + \frac{\displaystyle 1}{\displaystyle
            2}\left(2E - \frac{\displaystyle \mu}{\displaystyle
            R}\right)\right]^{2}}\right),
        \label{general-potential}\\
      V_{\infty} &=& 1 - \frac{1}{4}\left(2E + \frac{\mu}{R}\right)^2,
      \label{comoving-potential}
    \end{eqnarray}
        and $\lambda$ and $E$ are constants defined by $\lambda =
        \zeta_{+}/\xi$ and $E = M/\mu$ respectively. In the limit
        $\lambda \rightarrow \infty$, the potential $V$ coincide with
        $V_{\infty}$ in the comoving system. This represents the
        motion of the shell described by an observer corresponding to
        the time slicing parameter ``$\lambda$'' who is accelerated
        against gravity produced by the shell.

        It is worth noting that the metric component $N^2 =
        g_{tt}$ cannot be determined by the junction conditions. The
        reason can be seen in the proof of Birkhoff's
        theorem\cite{HawkingElis}. When one chooses $R$ to be the
        radial coordinate, the vacuum Einstein equations for the
        spherically symmetric vacuum spacetime give
    \begin{equation}
      \label{birkhoff}
      N^2 = \left(1 - \frac{2M}{R}\right)f(t)^2,
    \end{equation}
        where $f(t)$ is an arbitrary function of $t$. The factor
        $f(t)$ may be eliminated by using the coordinate
        transformation $dT = f(t)dt$. However, one cannot determine
        this function by the Einstein equations. Since the junction
        conditions are local conditions, all the metric components on
        the shell depend only on the time coordinate $t$, and the
        lapse function $N(t)$ is treated as an arbitrary function by
        the virtue of the gauge freedom $f(t)$ of a choice of $t$ by a
        local observer. In the next section, we develop the canonical
        quantum theory using this arbitrariness of $N$.

        Before discussing the quantization procedure, it is better to
        give the brief derivation of the classical solution of this
        system, since the vacuum spacetimes in the outer and inner
        regions of the shell are classically treated. In the comoving
        frame, the difference of the quasi-local mass between the
        outer and inner sides of $\Sigma$ leads to the formula
    \begin{equation}
        \label{global-mass}
      M = \frac{\mu}{2}(R'_{-} + R'_{+}),
    \end{equation}
        where prime denotes the derivative with respect to $\eta$.
        Then the junction conditions (\ref{nontrivial-junction}) allow
        us to write explicitly $R'_{-}$ and $R'_{+}$ as follows,
    \begin{equation}
        \label{r-dash}
      R'_{-} = \frac{1}{2}\left(2 E +
      \frac{\mu}{R}\right),\;\;\;\;\;\; R'_{+} = \frac{1}{2}\left(2 E
      - \frac{\mu}{R}\right).
    \end{equation}
        Because the classical motion for bound states is limited in
        the range,
    \begin{equation}
      0 \leq R \leq \frac{\mu}{2(1 - E)},
    \end{equation}
        the derivatives $R'_{+}$ and $R'_{-}$ must satisfy the
        condition
    \begin{equation}
      R'_{+} \leq 2 E - 1, \;\;\;\;\;\; R'_{-} \geq 1.
    \end{equation}
        The equalities hold just at the turning point $R=\mu/(2(1-E))$
        of the shell motion. Recall that $R'_{\pm}$ are proportional
        to the components of the extrinsic curvatures
        $K_{\theta\theta}^{\pm}$ of $\Sigma$ of the outer and inner
        sides. Hence, $K_{\theta\theta}^{-}$ is always positive, while
        $K_{\theta\theta}^{+}$ becomes negative when $E<1/2$. The
        signs of $K_{\theta\theta}^{\pm}$ are essential to the global
        structure of spacetime constructed by the junction of the
        outer and inner spacetimes. In the range of $E$ given by
        $0<E<1/2$ and $1/2<E<1$, the junction clearly shows a wormhole
        formation and a black hole formation, respectively (see
        Fig.\ref{fig:1})\cite{Blau}. The gravitational mass $M$ can be also
        negative, even if the local energy condition $\mu=\sigma
        R^{2}>0$ is imposed, and it does not contradict to the
        positive mass theorem \cite{positive-mass} because it has no
        asymptotically flat region and has a timelike singularity at
        the Schwarzschild side. Thus the relation between the value of
        $E$ and the global structure of spacetime is very clear, and
        we use this relation also in quantum mechanics of the shell,
        in which the motion is specified by a discrete eigenvalue of
        $E$.

    \begin{figure}[h]
      \begin{center}
        \leavevmode
        \epsfysize=7pc \epsfbox[158 208 633 383]{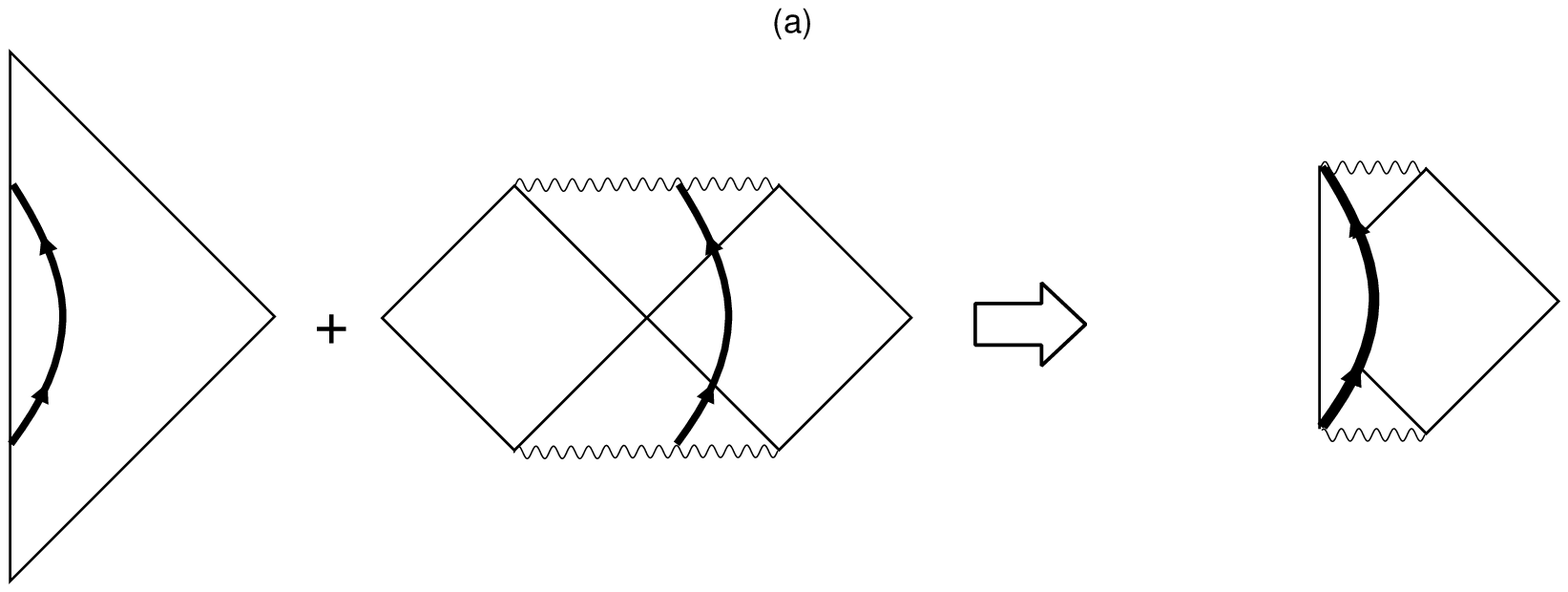}
      \end{center}
    \end{figure}
    \begin{figure}[h]
      \begin{center}
        \leavevmode
        \epsfysize=7pc \epsfbox[153 236 646 412]{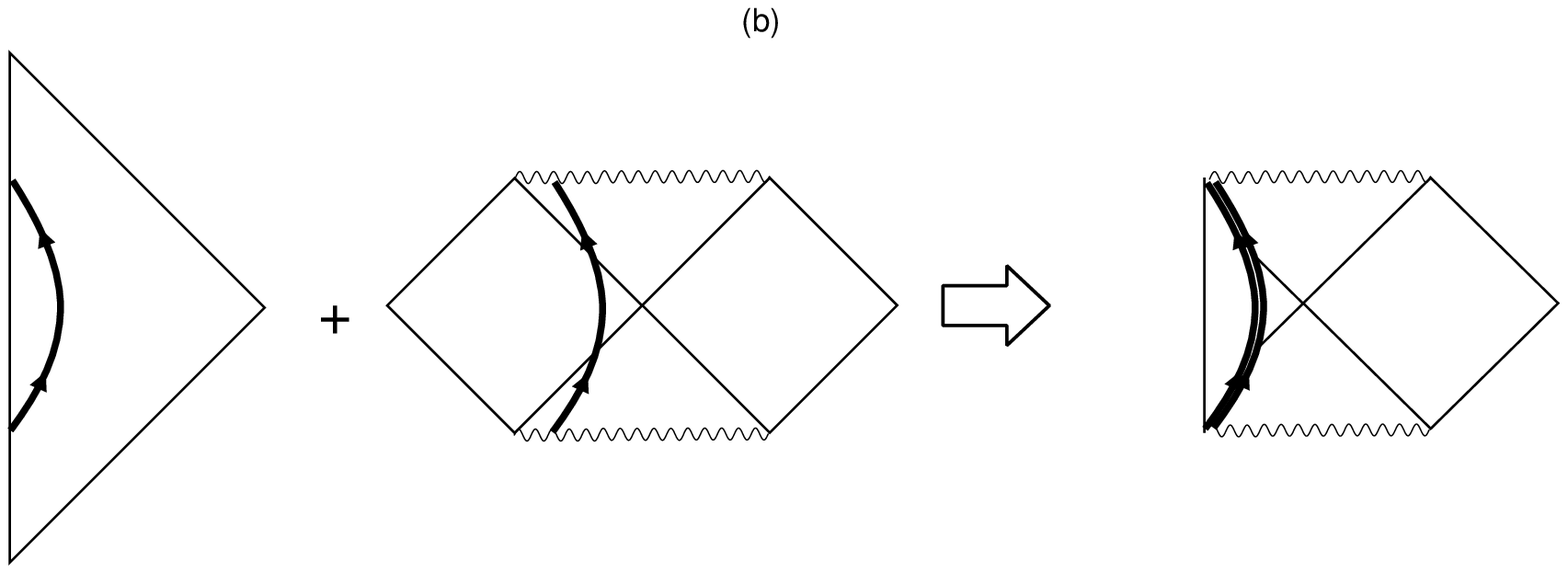}
      \end{center}
    \end{figure}
    \begin{figure}[h]
      \begin{center}
        \leavevmode
        \epsfysize=7pc \epsfbox[170 220 629 404]{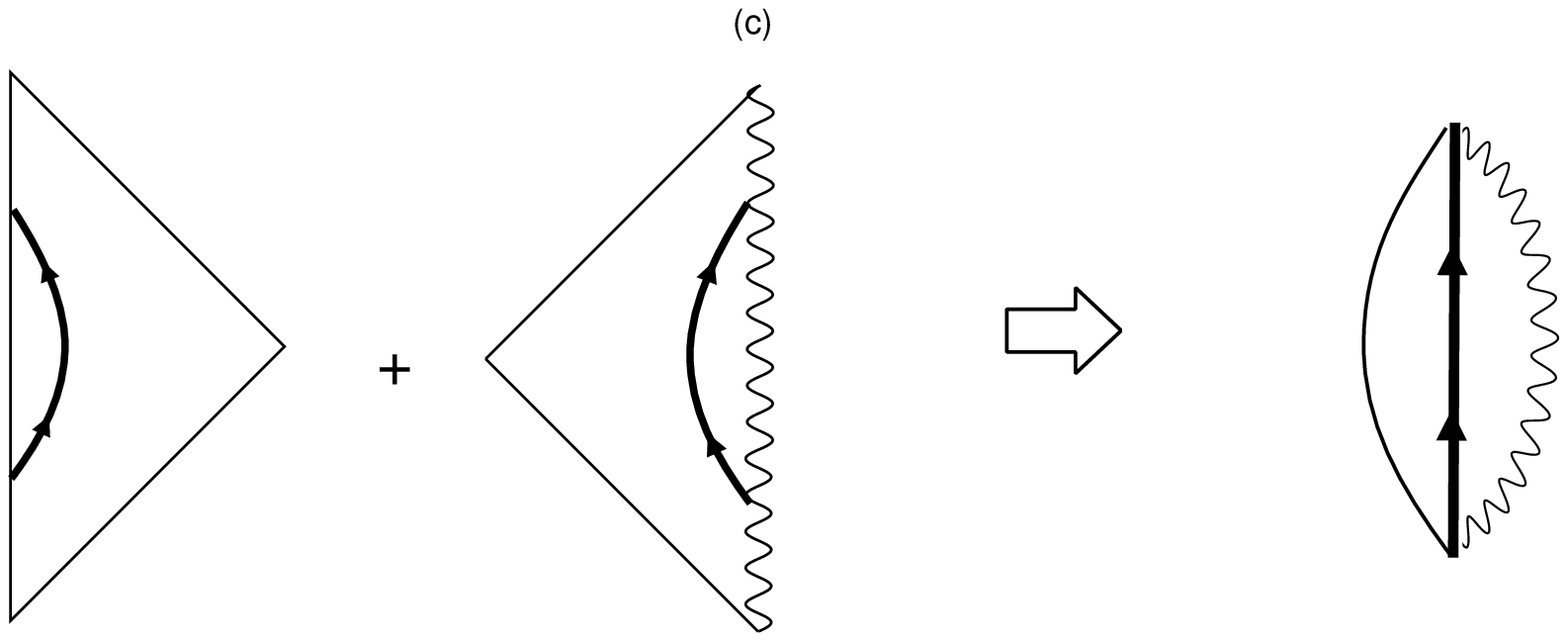}
      \end{center}
      \caption{Penrose diagrams of possible spacetimes constructed by
        the Minkowski-Schwarzschild junction with $E<1$. The
        trajectories of the dust shell in the Minkowski and
        Schwarzschild spacetimes are drawn. The two spacetimes are
        glued at the shell boundary. These figures show (a) a black
        hole spacetime for $1/2 < E < 1$ (in which the junction
        condition requires that both $R'_{+}$ and $R'_{-}$ are
        positive), (b) a wormhole spacetime for $0 < E < 1/2$ (in
        which $R'_{+} < 0$ and $R'_{-} > 0$), and (c) a negative mass
        spacetime (in which $R'_{+}<0$ and $R'_{-}$) without any
        asymptotic flat region. }
      \protect\label{fig:1}
    \end{figure}

\section{Canonical Quantization}
        \label{sec:canqua}

        Now we give the Hamiltonian which generates the equation of
        motion given by (\ref{general-equation-of-motion}). The
        procedure is to use the gauge freedom previously mentioned.
        Although the Hamiltonian is not uniquely determined, we seek
        the simplest one here. For this purpose, we consider the
        Lagrangian for (\ref{general-equation-of-motion})
    \begin{equation}
        \label{lagran}
      L =\frac{\mu}{2 N}\left(\frac{dR}{dt}\right)^2_{\eta} -
      \frac{N}{2 \mu}V.
    \end{equation}
        Because (\ref{lagran}) does not include $\dot{N}$, there is a
        primary constraint that is the canonical momentum conjugate to
        $N$ must weakly vanish. Furthermore, there is a secondary
        constraint which can be obtained by the variation of
        (\ref{lagran}) with respect to $N$. It is easy to confirm that
        the secondary constraint coincides with
        (\ref{general-equation-of-motion}). The Euler-Lagrange
        equation which can be derived by the variation with respect to
        $R$ is the first derivative of
        (\ref{general-equation-of-motion}). Because we have the
        Lagrangian (\ref{lagran}), the usual step of the canonical
        formalism leads to the canonical momentum conjugate to $R$
    \begin{equation}
      P = \frac{\mu}{N}\left(\frac{d R}{dt}\right)_{\eta},\\
    \end{equation}
        and the Hamiltonian
    \begin{equation}
        \label{hamiltonian}
      H = \frac{N}{2\mu}( P^2 + V ),
    \end{equation}
        We wish to emphasize that (\ref{general-equation-of-motion})
        is a result of the Hamiltonian constraint $H=0$ which is
        generated by the variation with respect to $N$. Our model can
        keep the property of a constrained system in a similar manner
        to the full canonical theory of general relativity.

        The next step is to quantize the Hamiltonian constraint given
        by
    \begin{equation}
      P^2 + V = 0.
          \label{hamicon-classical}
    \end{equation}
        If we use the usual commutation relation $[R,P]=i\hbar$ and
        the simplest factor ordering, (\ref{hamicon-classical}) is
        reduced to the Schr\"odinger equation,
    \begin{equation}
      \label{general-schrodinger}
      - \hbar^2 \frac{d^2}{dR^2} \Psi + V\Psi = 0,
    \end{equation}
        where potential $V$ is given by (\ref{general-potential}) and
        has the form
    \begin{equation}
      \label{general-potential-2}
      V = \frac{\mu^2[\lambda^2 + 1 + 2\lambda E](1 - E^2)}{(\lambda +
        E)^2}\frac{(R - R_{0})(R - R_{1})(R + R_{2})}{R(R - R_{3})^2},
    \end{equation}
        and the constants $R_{0}$, $R_{1}$, $R_{2}$ and $R_{3}$ are
        given by
    \begin{eqnarray}
      R_{0} &=& \frac{m(2E + \lambda) \hbar^{1/2}}{\lambda^2 + 1 + 2
        \lambda E}, \;\;\;\;\;\; R_{1} = \frac{m \hbar^{1/2}}{2(1 - E)},\\
      R_{2} &=& \frac{m \hbar^{1/2}}{2(1 + E)}, \;\;\;\;\;\; R_{3} =
      \frac{m \hbar^{1/2}}{2(\lambda + E)}.
    \end{eqnarray}
        Let us explain some implications of these radii. Because of
        the condition $|E|<1$ of the bound state, the classical motion
        of the shell has the maximum radius $R_{1}$ which corresponds
        to a turning point. In the time slicing parameterized by
        $\lambda$, the classical motion has also the minimum radius
        $R=R_{0}$ where the infinite redshift occurs for the
        corresponding observer. Note that $R_{0}$ is equal to $2M$ in
        the static limit $\lambda=0$, which coincides with the true
        horizon radius. From the potential
        (\ref{general-potential-2}), $R=R_{0}$ is a turning point of
        the shell. Since the WKB feature, in general, breaks down at
        turning points, this means that the semiclassical description
        become meaningless at $R=R_{0}$\cite{Hajicek-kouchan}. The
        potential $V$ diverges at $R=R_{3}$ where we obtain the
        regular singular point of the differential equation
        (\ref{general-schrodinger}). This radius $R_{3}$ becomes
        smaller than $R_{0}$, if the inequality $E>1/2-\lambda/2$
        holds. Then, we can consider the region $R_{3}<R<R_{0}$ in
        (\ref{general-schrodinger}), which is classically forbidden
        and disappears only in the comoving limit $\lambda \rightarrow
        \infty$ ({\it i.e.}, $R_{0} \rightarrow 0$). Only when
        $\lambda < 3$ there exist the additional bound states with $E$
        in the range $-1<E<1/2-\lambda/2$, for which we obtain the
        classically allowed region $R_{0} < R < R_{1}$ and classically
        forbidden regions $0 < R < R_{0}$ and $R_{1} < R < R_{3}$.
        This situation is similar to the quantum field theory in a
        Rindler spacetime, in which both sets of mode functions in
        the left- and right-handed wedges together are complete on all
        of Minkowski space\cite{Birrell-Davis}.

        The Schr\"odinger equation (\ref{general-schrodinger}) means
        that we have the Hilbert spaces ${\cal H}_{\lambda}$ of $\Psi$
        which depends on the time slicing parameter $\lambda$, in
        particular, as a consequence of the existence of the
        classically forbidden region $R_{3} < R < R_{0}$ (or $R_{0} <
        R < R_{3}$ when $-1 < E< 1/2 - \lambda/2$ and $\lambda < 3$).
        Hence, for each $\lambda$, we can give a discrete set of the
        eigenvalues of $E$, which is the unique observable in this
        quantum system. Because the vacuum spacetimes outside and
        inside the shell are classically treated, the global structure
        of the whole spacetime is specified only by $E$. Then the
        Hilbert space can be regarded as a set of spherically
        symmetric spacetimes (such as black holes, wormholes) which
        the collapsing shell forms.

\section{Mass Eigenvalues}
        \label{sec:eigen}

        In this section, we study the quantum mechanics on various
        time slicings using the Schr\"odinger equation
        (\ref{general-schrodinger}). We mainly consider the static
        time slicing $\lambda=0$ and then the other time slicing is
        considered to confirm that our argument is the natural
        extension of that in \cite{Hajicek-phys}.

\subsection{Static Time Slicing}

        First, we consider the typical time slicing which corresponds
        to a static observer. When $\lambda = 0$, the Schr\"odinger
        equation (\ref{general-schrodinger}) can be written in the
        form
   \begin{equation}
      \label{eqn:lam=0equation}
      -\hbar^2 \frac{d^2}{dR^2} \Psi + \frac{\mu^2(1 -
        E^2)}{E^2}\frac{(R - R_{0})(R - R_{1})}{(R - R_{3})^2}\frac{R
        + R_{2}}{R} \Psi = 0
    \end{equation}
        where
    \begin{eqnarray}
      R_{0} &=& 2M = 2 m E \hbar^{1/2}, \;\;\;\;\;\; R_{1} = \frac{m
        \hbar^{1/2}}{2(1 - E)},\nonumber\\
      R_{2} &=& \frac{m \hbar^{1/2}}{2(1 + E)},\;\;\;\;\;\; R_{3} =
      \frac{m \hbar^{1/2}}{2 E}.
    \end{eqnarray}
        As previously mentioned, for the bound states in the range
        $1/2<E<1$, only the bounded region $R_{0}\leq R\leq R_{1}$ is
        classically allowed. Though a quantum penetration of the wave
        function is possible in the region $R<R_{0}$, it must stop at
        $R=R_{3}$ owing to the infinite potential barrier. Therefore
        the boundary conditions which we adopt here is that the wave
        function vanishes at $R=R_{3}$ and $R\rightarrow\infty$.
        Although it is difficult to solve (\ref{eqn:lam=0equation})
        exactly, an approximate calculation of the eigenvalue $E$ is
        possible. Notice that the factor $1 + R_{2}/R$ satisfies the
        inequality,
    \begin{equation}
      1 < 1 - \frac{R_{2}}{R} < 1 - \frac{R_{2}}{R_{3}} = 1 +
      \frac{E}{1 + E},
    \end{equation}
        since the wave function is defined in the region $R_{3} < R <
        \infty$. As will be seen later, the eigenvalue $E$ is in the
        range $1/2<E<1$. Then, the factor
    \begin{equation}
      \label{black-hole-assumption}
      1< \alpha \equiv 1 + \frac{R_{2}}{R} < \frac{3}{2}
    \end{equation}
        remains nearly constant in (\ref{eqn:lam=0equation}). Then,
        introducing the non-dimensional variable $z$,
    \begin{equation}
      z = \frac{2 m \sqrt{\alpha(1 - E^2)}}{E}\frac{R -
        R_{3}}{\hbar^{1/2}},
    \end{equation}
         The approximate form of (\ref{eqn:lam=0equation}) can be
         written as follows,
    \begin{equation}
        \label{eqn:Whittaker}
      \frac{d^2}{dz^2}\Psi + \left(-\frac{1}{4} + \frac{k}{z} -
      \frac{p^2 - 1/4}{z^2}\right)\Psi = 0
    \end{equation}
        where $\alpha$ is treated as a constant, and
    \begin{eqnarray}
      p^2 &=& \frac{\alpha m^4(1 - 2E)^2 (1 + 2E)(1 + E)}{4 E^4} +
          \frac{1}{4},\\
      k &=& \frac{m^2(2E - 1)(2 + E - 2E^3)}{4E^2}\sqrt{\frac{(1 +
              E)\alpha}{1 - E}}
    \end{eqnarray}
        The general solution is a superposition of the two Wittaker's
        function $M_{k,p}(z)$ and $M_{k,-p}(z)$, one of which is
        defined by
    \begin{equation}
      \label{def-of-M}
       M_{k,p}(z) = z^{p+1/2} e^{-z}\sum^{\infty}_{n=0}
       \frac{\Gamma(2p + 1)\Gamma(p - k + n + 1/2)}{\Gamma(2p + n
         + 1)\Gamma(p - k + 1/2)} \frac{z^n}{n!}
    \end{equation}
        The boundary condition at $z \rightarrow \infty$ selects the
        unique solution $W_{k,p}(z)$ which exponentially decrease as
        $z$ increases and is written by the superposition
    \begin{equation}
       W_{k,p}(z)  =  \frac{\Gamma(-2p)}{\Gamma(1/2 - p
         -k)}M_{k,p}(z) + \frac{\Gamma(2p)}{\Gamma(1/2 + p
         -k)}M_{k,-p}(z).
    \end{equation}
        Now another boundary condition at the regular singular point
        $z=0$ ($R = R_{3}$) determines the eigenvalue $E$. The
        behavior $M_{k,p}(z) \rightarrow z^{p + 1/2}$ in the limit $z
        \rightarrow 0$ means that $\Psi$ can satisfy the boundary
        condition only when
    \begin{equation}
      \Gamma(1/2 + p - k) = \pm \infty,
    \end{equation}
        because the $\Gamma$-function does not vanish on the real axis.
        Since the $\Gamma$ function has poles at non-positive integers,
        this condition is reduced to
    \begin{equation}
      \label{spectrum-formula-schi}
      n = k - p - 1/2,
    \end{equation}
        where $n$ is a non-negative integer. From
        (\ref{spectrum-formula-schi}) we can give the limiting
        behavior of the mass eigenvalue: In the limit $n \gg m^2$,
    \begin{equation}
      E \sim 1 - \frac{\alpha m^4}{8 (n + 1)^{2}}.
          \label{eqn:nlargm-mass}
    \end{equation}
        If $\alpha=1$, (\ref{eqn:nlargm-mass}) corresponds to the
        spectrum (\ref{berezin-e}) obtained by Berezin\cite{Berezin90}
        and (\ref{comoving-eigen}) H\'aj\'\i\v{c}ek et
        al.\cite{Hajicek-phys} in the limit $n>>1$ of highly excited
        states. On the other hand in the opposite limit $n<<m^2$, the
        quantum effect becomes more important, and we have
    \begin{equation}
      E \sim \frac{1}{2} + \left(\frac{2n + 1}{\displaystyle \sqrt{3
          \alpha} m^2}\right)^{\frac{1}{3}}.
          \label{eqn:nsmalm-mass}
    \end{equation}

    \begin{figure}[h]
      \begin{center}
        \leavevmode
        \epsfysize=13pc \epsfbox[50 50 410 302]{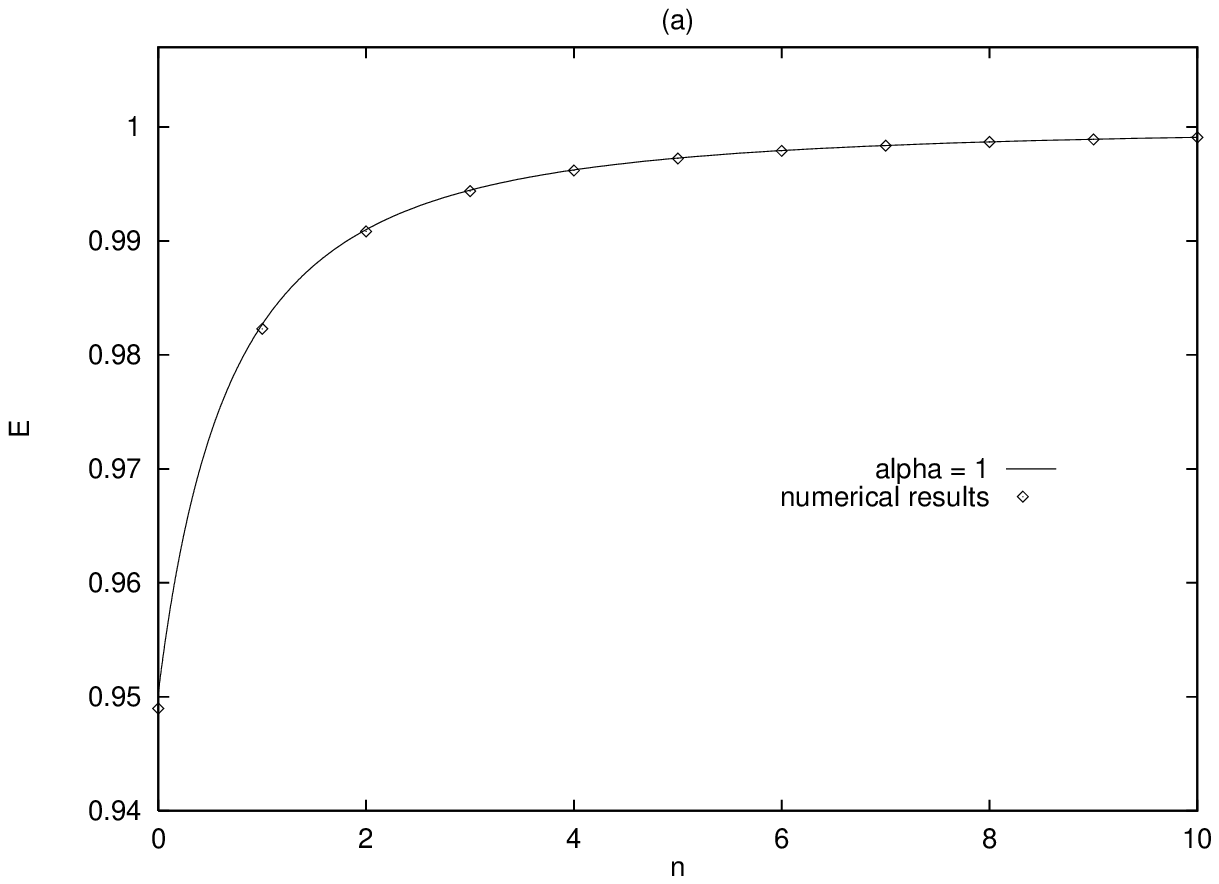}
      \end{center}
    \end{figure}
    \begin{figure}[h]
      \begin{center}
        \leavevmode
        \epsfysize=13pc \epsfbox[50 50 410 302]{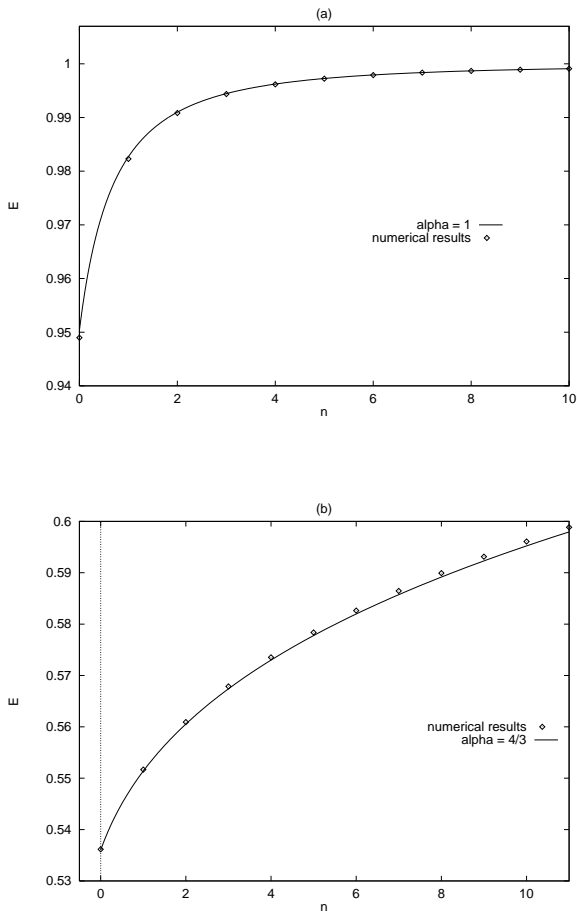}
      \end{center}
      \caption{The eigenvalues of $E$ for the wave functions of black
        hole formation in a static time slicing. The validity of the
        approximate formula (\protect\ref{spectrum-formula-schi})
        checked by numerical calculations. The value of $m$ is chosen
        to be (a) $m=1.0000$ and (b) $m=100.00$. The spectrum given by
        (\protect\ref{spectrum-formula-schi}) is drawn by solid lines
        which corresponds to (a) $\alpha=1$ and (b) $\alpha=4/3$
        respectively.}
      \protect\label{fig:2}
    \end{figure}

        Although the spectrum of $E$ obtained here is consistent with
        the assumption $1/2<E<1$, the validity can be also checked by
        numerical calculation of (\ref{eqn:lam=0equation}), which is
        based on the standard shooting method of solving two-point
        boundary-value problems\cite{Numerical}. In Fig.\ref{fig:2} the
        eigenvalues of $E$ are plotted for $m=1$ and $m=100$, which
        corresponds to the quantum numbers of $n$ which runs from $0$
        to $10$. We note that the approximate spectrum
        (\ref{spectrum-formula-schi}) coincides with the numerical
        results within the accuracy of our numerical code if choosing
        the constant $\alpha=1$ in (\ref{eqn:nlargm-mass}) and
        $\alpha=4/3$ in the case of
        (\ref{eqn:nsmalm-mass})\cite{alpha-4over3}. Hence we can claim that
        (\ref{spectrum-formula-schi}) is useful to discuss the
        qualitative behavior of the spectrum $E=E(n)$.

        The behaviors of the spectrum of $E$ written by
        (\ref{eqn:nlargm-mass}) and (\ref{eqn:nsmalm-mass}) are
        plausible as a quantum version of gravitational collapse.
        Recall that $n$ is the quantum number of the shell motion, and
        $1 - E$ can be regarded as the gravitational binding energy
        per unit rest mass energy. We can find that the gravitational
        binding energy remains small when the kinetic energy of the
        shell motion dominates the inertia of the shell ($n\gg m^2$),
        while the binding energy becomes large when the inertia
        dominates the kinetic energy. For the ground state ($n = 0$)
        we obtain $E\rightarrow 1/2$ in the limit $m\gg 1$, which
        corresponds to the classical limit of black hole
        formation. The quantum zero-point fluctuations generate the
        term $(2/m^2)^{1/3}$ in (\ref{eqn:nsmalm-mass}), where
        $\alpha$ is taken to be equal to $4/3$. The contribution of
        this zero-point fluctuations can seriously affect the motion
        of the shell, if $m$ is not so large.

        Note that the restriction $m<1$ which is required in the
        comoving time slicing disappears in this static time
        slicing. This is a consequence of the inner boundary
        condition. In this static time slicing, the central
        singularity is hidden by the infinite redshift surface, and
        the inner boundary condition is set up at a finite $R$. Hence
        we can construct quantum mechanics of the collapsing dust
        shell without taking account of ``information loss'' at the
        central singularity.

\begin{figure}[h]
      \begin{center}
        \leavevmode
        \epsfysize=13pc \epsfbox[50 50 410 302]{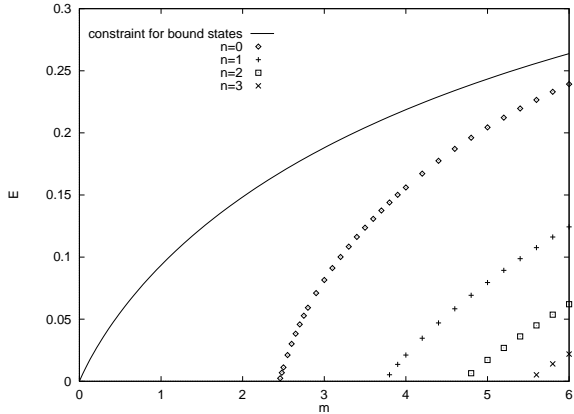}
      \end{center}
      \caption{The eigenvalues of $E$ for the wave functions of
        wormhole formation in a static time slicing. The
        $m$-dependence of $E$ is shown for the quantum numbers $n=0$,
        $1$, $2$ and $3$. The solid line given by
        $R_{3}>\pi/\protect\sqrt{-V_{\min}}$ means a rough upper
        boundary of the allowed range of $E$.}
      \protect\label{fig:3}
    \end{figure}

        The eigenvalues (\ref{spectrum-formula-schi}) (or
        (\ref{eqn:nlargm-mass}), (\ref{eqn:nsmalm-mass})) shows that
        the global structure of spacetime which corresponds to these
        eigenvalues is limited to only a black hole formation, {\it
          i.e.} the wave functions which has a support only in
        $R_{3}<R<\infty$ corresponds to the black hole
        states(Fig1.(a)). This does not means that there is no
        wormhole state. When $E<1/2$, we can also consider the wave
        function whose support exists only in $0<R<R_{3}$. In this
        case, the classically allowed region also exists in
        $R_{0}<R<R_{1}$ where the inequalities $0<R_{0}<R_{1}<R_{3}$
        holds due to the condition $E<1/2$. In this case, one cannot
        use the approximation that $\alpha$ defined by
        (\ref{black-hole-assumption}) is nearly constant. So, we must
        numerically solve the Schr\"odinger equation
        (\ref{eqn:lam=0equation}) and the result is shown in
        Fig.\ref{fig:3}: For fixed $m$, the eigenvalue of $E$
        monotonically decreases in the range $0<E<1/2$ as $n$
        increases. These eigenvalues plotted in Fig.\ref{fig:3}
        corresponds to the global structure of a wormhole formation
        {\it i.e.} the wave function whose support exists only in
        $0<R<R_{3}$ corresponds to the wormhole
        states(Fig.\ref{fig:1}(b)). When $m>>1$, there are many bound
        states whose eigenvalues of $E$ satisfy the inequality
        $E<1/2$. Together with the case $E>1/2$, we can take the limit
        to the positive mass classical solutions of bound state in
        which $E$ takes an arbitrary value in $0<E<1$. This situation
        is similar to the quantum field theory in Rindler
        spacetime. Note that there is no bound state in $\mu<m^{*}\sim
        2.4m_{p}$, and wormhole spacetimes do not exist in the case
        $\mu < m^{*}$ due to the zero point fluctuations of the shell
        motion.

        We must also note that the state of $E=1/2$ is forbidden like
        the case $E<1/2$. It can be also seen from the necessary
        condition for the existence of bound states. Let us denote the
        minimum value of the potential $V$ by $V_{min}(E)<0$ which
        depends on $E$. Since we impose the boundary condition
        $\Psi|_{R=0} = \Psi|_{R=R_{3}} = 0$, all eigenvalues $E$ of
        bound states must satisfy the condition
    \begin{equation}
      R_{3}(E) > \frac{\pi}{\sqrt{-V_{min}(E)}}.
    \end{equation}
        When $E=1/2-\delta, (\delta<<1)$, we obtain $V_{min}(E) \sim -
        m^2 \delta^{2}(1 + \delta)$. Then the inequality is reduced to
        $m^2 > \pi/\delta$, which cannot be satisfied when
        $\delta\rightarrow 0$. Furthermore we can give a physical
        interpretation of $E\neq 1/2$. The classically allowed region
        is $R_{0}<R<R_{1}$, and $R_{1}-R_{0}\rightarrow 0$ as
        $E\rightarrow 1/2$, so the shell is confined in this narrow
        region. However, it is impossible due to the uncertainty
        relation $\Delta P\cdot\Delta R\sim\hbar$, thus $E=1/2$ is
        forbidden due to the quantum effect of the shell motion.

\subsection{Non-static Time Slicing}

\begin{figure}[h]
      \begin{center}
        \leavevmode
        \epsfysize=13pc \epsfbox[50 50 410 302]{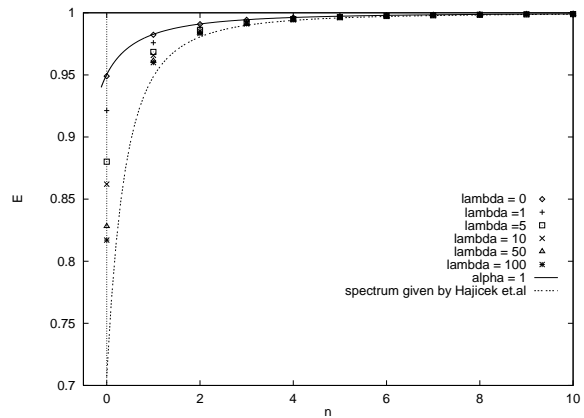}
      \end{center}
      \caption{The $\lambda$-dependence of the spectrum $E(n)$ for
        $m=1.0000$. The time slicing parameter is chosen to be
        $\lambda=0.0000$, $1.0000$, $5.0000$, $10.000$, $50.000$ and
        $100.00$. As $\lambda$ increases, the spectrum of $E$
        approaches to (\protect\ref{comoving-eigen}) drawn by the
        dashed line.}
      \protect\label{fig:4}
    \end{figure}

\begin{figure}[h]
      \begin{center}
        \leavevmode
        \epsfysize=13pc \epsfbox[50 50 410 302]{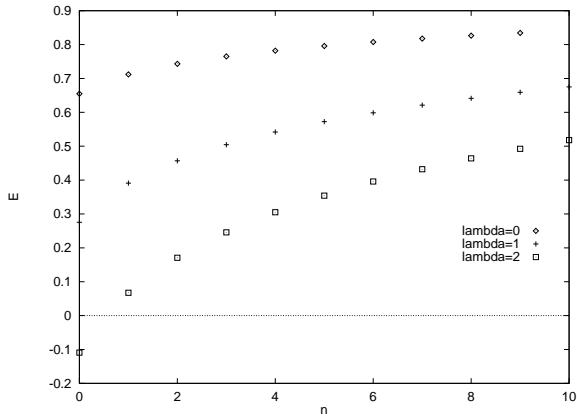}
      \end{center}
      \caption{The $\lambda$-dependence of the spectrum $E(n)$ for
        $m=10.000$. The time slicing parameter is chosen to be
        $\lambda=0.0000$, $1.0000$ and $2.0000$. It is shown in this
        figure that the spectrum does not remain in the range
        $1>E>1/2$ and even the negative mass states are allowed.}
      \protect\label{fig:5}
    \end{figure}

        Based on the result obtained in the static time slicing, we
        consider the Schr\"odinger equation
        (\ref{general-schrodinger}) in the case $\lambda \neq 0$
        except that there is no bound states in the region $0<R<R_{3}$
        when $\lambda>3$. Any essential property of
        (\ref{general-schrodinger}) is not so much different from the
        case $\lambda = 0$, and we also impose the boundary condition;
        the wave function must vanish at the regular singular points
        $R = m \hbar^{1/2}/(2 ( \lambda + E ))$ and $R = \infty$.
        Then, we can derive the eigenvalues $E$ in a similar manner.
        The approximation that $\alpha = 1 - R_{2}/R$ is constant in
        (\ref{general-schrodinger}) leads to
    \begin{equation}
      E \sim 1 - \frac{\alpha m^4 (1 + \lambda)^2}{8 (n + 1)^{2}}.
          \label{eqn:n0lambda-nlargm-mass}
    \end{equation}
        in the limit $n >> m^2$. This shows that the
        $\lambda$-dependence of $E$ is not so sensitive in the limit
        $n>>m^2$. However, the approximated form
        (\ref{eqn:n0lambda-nlargm-mass}) will not remain valid as
        $\lambda$ become infinitely large. To clarify the behavior of
        $E$ for the ground states in the limit $\lambda >> 1$, we must
        solve numerically the Schr\"odinger equation
        (\ref{general-schrodinger}). By varying the parameter
        $\lambda$, we can consider the extrapolation from the static
        time slicing to the comoving one. In particular, for $m\leq
        1$, we can compare the spectrum of $E$ with
        (\ref{comoving-eigen}) in the comoving limit. The numerical
        results for $m=1$ are plotted in Fig.\ref{fig:4}, which
        confirm that the mass spectrum converges to
        (\ref{comoving-eigen}) as $\lambda$ increases and $E$ remains
        larger than $1/2$. On the other hand, as an example of the
        spectrum for $m>1$, the eigenvalues $E$ for $m=10$ are plotted
        in Fig.\ref{fig:5}. We find the common tendency that the
        energy level at the ground state decreases as $\lambda$
        increases. The remarkable point for $m>1$ ({\it i.e.}, $\mu >
        m_{p}$) is that the mass spectrum does not keep the condition
        of $E>1/2$ nevertheless the wave function has its support only
        in $R_{3} < R < \infty$. As was suggested in
        \cite{Hajicek-phys}, even the states of $E<0$ are allowed, if
        $\lambda$ is sufficiently large. (This might means that
        $H_{B}$ in (\ref{B-h}) cannot be positive self-adjoint
        operator in general.) Then observable states exist in the
        range $-1 < E < 1$ when $\mu$ is sufficiently larger than
        $m_{p}$, and these correspond to the global structure of black
        hole and wormhole formations (see Fig.\ref{fig:1}).


        Finally, we consider the wave function given by the
        Schr\"odinger equation (\ref{general-schrodinger}) in the
        comoving frame.($\lambda \rightarrow \infty$), which has the
        form
    \begin{equation}
      - \hbar^2 \frac{d^2}{dR^2}\Psi + \mu^2 \left(1 - \frac{1}{4}
      \left( 2E + \frac{\mu}{R}\right)^{2}\right)\Psi = 0.
          \label{comoving-eq}
    \end{equation}
        Note that this Schr\"odinger equation corresponds to the
        Wheeler-DeWitt equation (\ref{Hajicek-WD}), and our
        Schr\"odinger equation may be regarded as a natural extension
        of (\ref{Hajicek-WD}). In terms of the requirement of
        unitarity and of positivity of energy, H\'aj\'\i\v{c}ek et
        al\cite{Hajicek-phys} gave the boundary condition that for the
        bound state $E < 1$ the wave function must vanish at the
        origin and at infinity. Then the discrete spectrum of $E$ was
        shown to be (\ref{comoving-eigen}) which becomes meaningless
        when $m > 1$. As discussed by H\'aj\'\i\v{c}ek et al., this
        means that when $m>1$ one cannot impose the positivity of $E$
        or the regularity of the wave function at $R=0$. In our
        approach the regularity of wave function at $R = R_{3}$ are
        set up in any time slicing parameter $\lambda$, even if
        $m>1$. Therefore, we can discuss the comoving limit through
        the extrapolation of the results to the range $\lambda
        \gg1$. In this sense, as previously mentioned, observable
        states exist in the range $-1 < E < 1$ when $\mu$ is
        sufficiently larger than $m_{p}$, there is no essential
        differences in a set of observable states. On the other hand,
        there are no wormhole states $0 < E < 1/2$ when
        $\mu<m_{p}$. These states are suppressed by the zero point
        fluctuations of the shell motion.

\section{Summary and Discussion}
  \label{sec:sum}

        In summary, by studying this model of dust shell collapse, we
        obtained the wave function for the bound states with positive
        eigenvalues of $E$ under the static time slicing. When $\mu
        >> m_{p}$, these can recover the classical solutions which
        describe wormhole or black hole formation. One may regard the
        regions $0<R<R_{3}$ and $R_{3}<R<\infty$ as left- and right-
        handed wedge of Rindler spacetime respectively in a similar
        way to the quantum field theory in Rindler
        spacetime\cite{Birrell-Davis}: The set of classical solution
        with positive eigenvalues of $E$ in our model corresponds to
        the complete set of mode functions in whole Minkowski
        spacetime, and black hole and wormhole states correspond to
        the mode functions in left- and right-handed wedge
        respectively. Furthermore, we showed that the wormhole states
        cannot exist when $\mu < m^{*} \sim 2.4 m_{p}$ as a result of
        the zero point fluctuations of the shell motion. For a
        comoving time slicing also, we can see the same quantum
        effect. Although it is not unclear whether $E<0$ is
        permissible or not, observable states exist in the range $-1 <
        E < 1$ when $\mu > m_{p}$, while only black hole states are
        possible when $\mu<m_{p}$. Thus, our conclusion is that the
        wormhole formation with small mass is efficiently suppressed
        in the quantum collapse of dust shell. Then, in our model, it
        is difficult to create small wormhole due to the quantum
        fluctuations of the shell motion.

        Although the $\lambda$-dependence of $E$ is not so essential
        to the set of observalbe states in our argument, to set a time
        slicing, we can easily see the relation between observalbe
        states and geometries (wormhole states or black hole states,
        etc.). This means that the time slicing or observers in a
        spacetime plays an important role when one classify the
        observable states in quantum theory of gravity. Since quantum
        theories are based on the Cauchy problem and, usually, the
        configuration space of quantum gravity is three-space
        geometries, we must specify a set of observers or initial time
        slicing at first to clarify the set of observable states. In
        this sence, the notion of the time slicing or observer will
        become to be more important in a fully consistent quantum
        gravity.

\acknowledgments

        We would like to thank H.Ishihara, J.Soda, T.Tanaka and
        M.Hotta for valuable discussions.


\begin{references}
\bibitem{alpha-4over3}
  $\alpha=4/3$ can be easily obtained by the estimation as
  follows. When $E \rightarrow 1/2$, the classically allowed region
  become to be narrow due to the $R_{1}-R_{0} \rightarrow 0$. Since
  the eigenvalues of $E$ is determined by the behavior of the wave
  function in the classically allowed region, $\alpha \sim 1 +
  R_{2}/R_{0} \sim 1 + R_{2}/R_{1} \sim 4/3$ when $E \rightarrow
  1/2$.
\end{references}
\end{document}